\documentclass[preprint,showpacs,preprintnumbers,amsmath,amssymb]{revtex4}
\usepackage{bm}% bold math
\usepackage{epsf}

\everymath{\displaystyle}
\begin{document}           % End of preamble and beginning of text.
%\pagestyle{empty}
%%%%%%%%%%%%%%%%%%%%%%%%%%%%%%%%%%%%%%%%%%%%%%%%%%%%%%%%%%%%%
\title{Effective electric and magnetic polarizabilities of pointlike spin-1/2 particles}

\author{A. J. Silenko}\email{alsilenko@mail.ru}
\affiliation{Bogoliubov Laboratory of Theoretical Physics, Joint Institute for Nuclear Research,
Dubna 141980, Russia \\ Research Institute for Nuclear Problems, Belarusian State University, Minsk 220030, Belarus}

\begin{abstract}
Effective electric and magnetic polarizabilities of pointlike spin-1/2
particles possesing an anomalous magnetic moment are calculated
with the transformation of an initial Hamiltonian to the
Foldy-Wouthuysen representation. Polarizabilities of spin-1/2 and spin-1 particles are compared.
\end{abstract}

 \maketitle

%%%\section{Введение}

In this work, the
Foldy-Wouthuysen (FW) transformation is used to determine the electric and magnetic polarizabilities of pointlike spin-1/2 particles
possessing an anomalous magnetic moment (AMM).

The unique properties of the FW representation \cite{FW} make it a very convenient tool for transition to semiclassical approximation and finding the classical limit
of relativistic quantum mechanics. Even for relativistic
particles in an external field, the operators in this representation are completely analogous to the corresponding operators of nonrelativistic quantum
mechanics. In particular, the localization operators
(the Newton-Wigner operators) and the momentum
operators are equal to \cite{NW} $\bm r$
and $\bm p=-i\hbar\nabla$, and the polarization operator for the spin-1/2 particles is the Dirac matrix $\bm \Pi$. In other representations,
these operators are defined by considerably more
cumbersome formulas (see \cite{FW,JMP}). Apart from the
simple and unambiguous form of operators that correspond to the classical observed ones, the most significant merit of FW representation is the restoration of
the probabilistic interpretation of the wave function.
Since, as was said above, it is in the FW representation
that the Newton-Wigner operator, which characterizes the location of a particle’s geometric center, is
equal to the radius-vector $\bm r$, the squared wave function
modulus determines the probability density of the particle’s location at the point with the given radius vector. We should note that, in the FW representation, the
Hamiltonian and all operators are diagonal over two
spinors (block-diagonal). Using this representation
eliminates the possibility of ambiguous solutions for
the problem of finding the classical limit of relativistic
quantum mechanics \cite{FW,CMcK}.

The most essential parameters that characterize the
particles and the nuclei are scalar electric and magnetic polarizabilities. Their contribution to the Hamiltonian in the FW representation is determined by the
expression
\begin{equation}\begin{array}{c}
\Delta {\cal H}_{FW}=-\frac{1}{2}\alpha_sE^2-\frac{1}{2}\beta_sB^2.\end{array} \label{opr}\end{equation} These parameters are \emph{effective} polarizabilities which origin is purely quantum mechanical. In particular, any real separation of electric charges of pointlike particles do not take place.

We consider here a case of stationary and uniform electric ($\bm E$) and magnetic ($\bm B$) fields and use the
system of units $\hbar=1,~c=1$.

While in the classical physics a particle may have
nonzero polarizabilities only provided that it has an
internal structure, in quantum mechanics even pointlike objects appear to have nonzero polarizabilities.
These parameters can be defined by the FW transformation of the initial Dirac–Pauli equation \cite{P} for particles with an AMM followed by transition to the classical limit. Solving this problem necessitates, however,
the calculation of external-field quadratic summands.
This procedure requires caution, since different techniques yield different results (see
\cite{dVFor} and references therein). Correct results are obtained by the
Eriksen method \cite{E}. It is convenient to divide the initial Dirac–Pauli Hamiltonian \cite{P} into the even and 
odd summands that commutate and anticommutate
with the Dirac matrix $\beta$, respectively:
\begin{equation} {\cal H}_D=\beta m+{\cal E}+{\cal
O}, ~~~\beta{\cal E}={\cal E}\beta, ~~~\beta{\cal O}=-{\cal
O}\beta. \label{eq3Dirac} \end{equation}
Here,
\begin{equation} {\cal E}=e\Phi-\mu'\bm
{\Pi}\cdot \bm{B}, ~~~{\cal O}=c\bm{\alpha}\cdot\bm{\pi}+i\mu'\bm{\gamma}\cdot\bm{ E}, \label{DireqIII} \end{equation}
where $\mu'$ is the AMM. We use the conventional denotations \cite{BLP} for the Dirac matrices.

The expansion in $1/m$
powers of the Hamiltonian
in the FW representation obtained by the Eriksen
method is presented in Refs.
\cite{dVFor,TMPFW}. In the case under
consideration, the summands proportional to the
fourth and higher powers of the reciprocal mass can be neglected. In this case, the Hamiltonian in the FW
representation is defined according to the equation
\begin{equation}\begin{array}{c}
{\cal H}_{FW}=\beta\left(m+\frac{{\cal O}^2}{2m}-\frac{{\cal
O}^4}{8m^3}\right)+  {\cal E}-\frac{1}{8m^2}[{\cal O},[{\cal O},
{\cal F}]]+\frac{\beta}{16m^3}\left\{{\cal O},\left[[{\cal O},{\cal
F}],{\cal F}\right]\right\},
\end{array} \label{ProE1}\end{equation}
where ${\cal F}={\cal E}-i\partial/\partial t$. For the considered stationary problem, ${\cal F}={\cal E}$.

Calculation according to Eqs. (\ref{DireqIII}) and (\ref{ProE1}) yields the
following expression:
\begin{equation}\begin{array}{c} {\cal H}_{FW}=\beta\left(m+
\frac {\bm \pi^2}{2m}-\frac{\bm \pi^4}{8m^3}\right)+ e\Phi+\frac{
  1}{2m}\left(\frac{
   \mu_0}{2}+\mu'\right)\left(2\bm\Sigma\cdot[\bm\pi\times\bm
E]-\nabla\cdot\bm E\right)\\-\left(
   \mu_0+\mu'\right)\bm\Pi\cdot\bm B+\frac{
   \mu'}{4m^2}\left\{\bm\Pi\cdot\bm\pi,\bm\pi\cdot\bm B\right\}+\beta\frac{(\mu_0+\mu')\mu'}{2m}E^2-\beta\frac{
   \mu_0^2}{2m}B^2,
\end{array} \label{eq33new} \end{equation}
where $\mu_0=e/(2m)$ is the Dirac magnetic moment.

We should note that the last summand in Eq. (\ref{ProE1}) does not make any contribution to the electric and
magnetic polarizabilities. Calculations performed by the method
developed by Foldy and Wouthuysen \cite{FW} and by other iterative methods (see Refs. 
\cite{TMPFW,JMPcond} and references therein) lead to a different form of this
summand and, as a consequence, do not yield the correct expression for the electric polarizability. A comparison of Eqs. (\ref{opr}) and (\ref{eq33new}) shows that the scalar electric
and magnetic polarizabilities of the pointlike particles
possessing an AMM have the form
\begin{equation} \begin{array}{c}
\alpha_S=-\frac{(\mu_0+\mu')\mu'}{m}=-\frac{e^2g(g-2)}{16m^3}, ~~~
\beta_S=\frac{\mu_0^2}{m}=\frac{e^2}{4m^3}.
\end{array} \label{eqplrsh} \end{equation}
The matrix $\beta$ can be omitted since, in the FW representation, the lower spinor is equal to zero.

A comparison of polarizabilities of pointlike spin-1/2 and spin-1 particles is important. The spin-1 particles
are characterized by not only scalar but also tensor
polarizabilities which are calculated in Ref. \cite{PRDspin1}. The scalar polarizabilities of such particles are defined by the expressions \cite{PRDspin1}:
\begin{equation} \begin{array}{c}
\alpha_S=-\frac{e^2(g-1)^2}{m^3}, ~~~ \beta_S=0.
\end{array} \label{eqplriz} \end{equation}
Thus, the scalar magnetic polarizability of the spin-1
particles equals zero and the scalar electric polarizability is nonzero for particles with not only an anomalous but also normal $(g=2)$ magnetic moment. These
properties differ from the corresponding properties of
the spin-1/2 particles.

We should note that the scalar polarizabilities of
pointlike spin-0 particles are equal to zero (see Ref. \cite{TMP2008}).

The scalar polarizabilities of pointlike particles belong to quantum mechanical effects. Such particles do not have any charge distribution and their interaction with the electric field does not lead to any charge separation. 

\medskip 
The work was supported by the Belarusian Republican Foundation for
Fundamental Research (Grant No. $\Phi$14D-007).

\end{document}